\documentclass[11pt]{article}
\usepackage[english]{babel}

\usepackage{graphicx}
\usepackage{amsmath}
\usepackage{amssymb}
\usepackage{amsxtra}
\usepackage{mathrsfs}

\usepackage[backend=biber,style=gost-numeric,autolang=other,bibencoding = auto,sorting = none,doi = false,url = false]{biblatex}
\usepackage{csquotes}
\title{The interaction of domain walls with fermions in the early Universe}
\author{Andrey A. Kurakin$^1$\\e-mail kurakin-1993@mail.ru\\Sergey G. Rubin$^{1,2}$\\ e-mail sergeirubin@list.ru\\\\
$^1$ National Research Nuclear University MEPHI\\
(Moscow Engineering Physics Institute),\\
$^2$ N.I. Lobachevsky Institute of Mathematics and Mechanics,\\ Kazan  Federal  University,  {Kremlevskaya  Street  18}, \\ 420008 Kazan,  Russia
115409 Moscow, Russia}

\begin{document}
\maketitle

\begin{abstract}
We consider the scalar field solitons and their interaction with the fermions in the early Universe.
The analytical form of the reflection coefficient is obtained. The fermion mass is a function of the distance between the fermion and the soliton (wall). The function was approximated by the Woods-Saxon potential.
\end{abstract}



\section{Introduction}\label{s:intro}

Primordial black holes (PBHs) have been a source of significant interest for over 50 years. The possibility of the existence of such objects was predicted by Zeldovich and Novikov \cite{zel1967hypothesis}. Despite the absence of direct evidence of their existence, there is a lot of observational data that can be interpreted in the framework of the hypothesis of the origin of black holes (BH) at the initial stages of the origin of the Universe \cite{Clesse:2017bsw, Ali-Haimoud:2019khd, Carr:2020xqk}.

In this paper, we base on the model of PBH formation as a result of the collapse of domain walls \cite{Rubin, Belotsky:2018wph, Deng:2018cxb}. As a result of phase transitions during and after the inflationary stage, closed domain walls are formed. The formed non-spherical wall evolves: when interacting with hot plasma, the kinetic energy of the wall dissipates. As a result, the oscillations of the domain wall decay, the energy is transferred to the surrounding plasma, which leads to its additional heating. Further, the wall spheres and collapses into BH.

The rate of energy transfer from the domain wall to the surrounding plasma depends on the wall thickness, the initial plasma temperature and its density. The wall thickness is characterized by the parameters of the initial Lagrangian and can vary over a wide range. Plasma temperature and density depend on the moment the walls appear. Moreover, the dynamics of plasma parameters depends on whether it participates in cosmological expansion or is separated from it due to the gravitational well created by closed walls.

In this paper, we consider the fermion interaction with the scalar field solitons (walls).

\section{Model of domain wall}\label{s:model}

Consider the domain wall model. We describe We describe the wall by a complex scalar field with a Lagrangian:
\begin{equation}\label{eq:lagrwall}
    \mathcal{L}_{wall}=\partial_{\mu}\phi^{\dag}\partial^{\mu}\phi - \frac{1}{4}(\phi^{\dag} \phi - f^2/2)^2 - \Lambda^4(1-\cos \theta),
\end{equation}
where $\phi$ - complex scalar  field and $\theta$ is its phase.
At the end of inflation, the $\phi$ field is captured by the potential minimum for which $|\phi|=f$. Then we write the complex field in the form:
\begin{equation}\label{eq:phi}
    \phi = \frac{f}{\sqrt{2}}e^{i\theta} =\frac{f}{\sqrt{2}}e^{i\chi/f}.    
\end{equation}
Substitution of the expression \eqref{eq:phi} into Lagrangian \eqref{eq:lagrwall}
 gives Lagrangian, that describing the phase of complex scalar field: 
 \begin{equation}\label{eq:lagrwall2}
    \mathcal{L}_{wall}=\frac{1}{2}(\partial_{\mu}\chi)^2 - \Lambda^4(1-\cos(\chi/f) ).
\end{equation}
The the phase $\chi$ is determined as follows \cite{Rajaraman:1982is}:

\begin{equation}\label{eq:chi}
\chi (x) = 4f\arctan\left(\exp\left[\frac{\Lambda^2}{f}x\right]\right)=4f\arctan\left(\exp\left[\frac{2x}{d}\right]\right),
\end{equation}
where we introduced the wall thickness parameter $d$
\begin{equation}\label{eq:d}
d = \frac{2f}{\Lambda^2}.
\end{equation}

Let us choose the Lagrangian of fermions in the form:
\begin{equation}\label{eq:lagrferm}
\begin{aligned}
    \mathcal{L}_{f}=i\bar\psi \gamma^{\mu}\partial_{\mu}\psi + g_0(\phi\bar\psi \psi + h.c.) - m\bar\psi\psi =\\ =i\bar\psi \gamma^{\mu}\partial_{\mu}\psi + \sqrt{2}g_0f\bar\psi \psi\cos(\chi/f) - m\bar\psi\psi.
\end{aligned}
\end{equation}
where expression \eqref{eq:phi} is used.

The interaction of the fermions with the domain wall is
\begin{equation}
     \mathcal{L}_{int} = m_0 \cos(\chi/f)\bar\psi\psi =  m_0\left(1-\frac{2}{\cosh^2(2x/d)} \right)\bar\psi\psi; \quad  m_0=\sqrt{2}fg_0 .
\end{equation}\label{eq:lagrint2}
Then Lagrangian of fermions can be rewritten as
\begin{equation}\label{eq:lagrferm2}
    \mathcal{L}_{f}=i\bar\psi \gamma^{\mu}\partial_{\mu}\psi -m_0\frac{2}{\cosh^2(2x/d)}\bar\psi \psi -m_f\bar\psi\psi, 
\end{equation}
where $m_f=m-m_0$ - fermion mass.

\section{Dirac equation}\label{s:equation}

A description of the interaction between  fermions and  domain wall within the framework of the approach to solving the equation of motion is given in the papers \cite{Ayala:1993mk,PhysRevD.73.105021,Farrar:1994vp}
The result for the interaction of the wall with scalar particles is given in the monograph \cite{Vilenkin:2000jqa}. In the papers \cite{Ayala:1993mk,PhysRevD.73.105021,Farrar:1994vp} the description of the domain wall is given by the kink solution:  $\phi \sim \tanh x$. In such model, the asymptotic fermion mass takes different values: $x \rightarrow \pm \infty$. 
This problem does not arise for the Lagrangian  \eqref{eq:lagrferm2}: the fermion mass is the same on both sides of the wall.

The Dirac equation 
\begin{equation}
    0 = \left(i\gamma^{\mu}\partial_{\mu} - g(x)\right)\psi,
\end{equation}
holds for fermion Lagrangian \eqref{eq:lagrferm2}
where function
\begin{equation}\label{eq:lagrint2}
    g(x) = \frac{2m_0}{\cosh^2 (2x/d)} + m_f
\end{equation}
is effective mass, depending on the coordinate in the  coordinate $x$ perpendicular to the wall. Hereinafter, in asymptotics, we have: $g(x) \xrightarrow{x \rightarrow \pm \infty} m_f$.

The fermion wave function is as follows
\begin{equation}
    \psi(x) = \begin{pmatrix}
    u_1(x) & u_2(x) & u_3(x) & u_4(x)
    \end{pmatrix}^{T}e^{-iEt+ip_tx_t}.
\end{equation}
Here we put $p_t=0$ for simplicity, i.e. the component of the momentum  in the plane of the domain wall is zero and the incident wave is perpendicular to the wall. Then the equation takes the form:
\begin{equation}
    0 = \left(E\gamma^0+i\gamma^{3}\partial_x - g(x)\right)\psi(x).
\end{equation}
Hereafter, we choose the following representation of gamma matrices:
\begin{equation}
    \gamma^0 = \begin{pmatrix}
    0 & 1 \\ 1 & 0
    \end{pmatrix}, \;\; 
    \gamma^3 = \begin{pmatrix}
    0 & \sigma^3 \\ -\sigma^3 & 0
    \end{pmatrix}
\end{equation}
As a result of substitution, we obtain a system of equations for the bispinor components:
\begin{equation}
\begin{aligned}
  0 =Eu_3(x) + iu'_3(x) - g(x)u_1(x) \\
  0 =Eu_1(x) - iu'_1(x)- g(x)u_3(x) .
\end{aligned}
\end{equation}
We obtain a similar system for the components $u_2,\; u_4$ if we replace: $u_1 \rightarrow u_4,\;u_3 \rightarrow u_2$.

Let's consider the following linear combinations of bispinor components:
\begin{equation}
\begin{aligned}
    \phi_+(x) = u_1(x) + iu_3(x) \\
    \phi_-(x) = u_1(x) - iu_3(x).
\end{aligned}    
\end{equation}
As a result of such substitution we obtain a system:
\begin{equation}\label{eq:syst}
\begin{aligned}
    0 = iE\phi_-(x) + \phi'_+(x) - g(x)\phi_+(x) \\
    0 = iE\phi_+(x) + \phi'_-(x) + g(x)\phi_-(x).
\end{aligned}    
\end{equation}
Excluding the variables, we obtain the equations for the components $\phi_{\pm}(x)$:
\begin{equation}\label{eq:eq}
\begin{aligned}
    0 = \left(\frac{d^2}{dx^2} \mp g'(x)+E^2-g^2(x) \right)\phi_{\pm}(x). \\
\end{aligned}    
\end{equation}

 Let us carry out an approximation by a function for which the solution can be obtained in an analytical form. Let us choose the Woods-Saxon potential. The scattering problem for the Woods-Saxon potential is considered in detail in the papers \cite{Kennedy:2001py, Panella:2010yb}:
After approximation, the function $g(x)$ takes the form
\begin{equation}
    g(x) = \frac{A\theta(x)}{1+exp(a(x-x_0))} +   \frac{A\theta(-x)}{1+exp(-a(x+x_0))} + m_f,
\end{equation}
where parameters: $A \rightarrow A = 2.392m_0$, $m_f = m - m_0$.

\begin{figure}[h!]\center
	\includegraphics[scale=0.7]{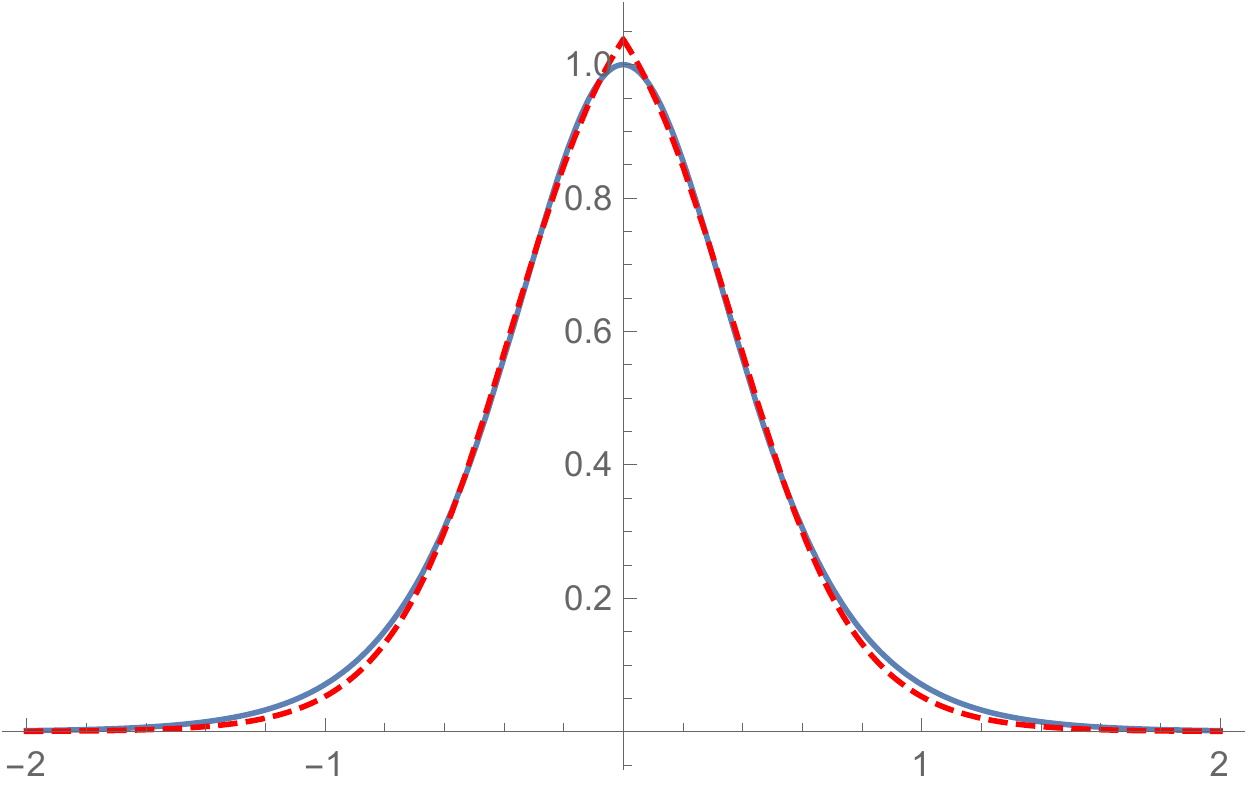}
		\caption{Approximation of the function g (x). Blue solid line - g (x), red dotted line - Woods-Saxon potential}
	\end{figure}
	
We solve the equation for two regions: $x<0$ and $x>0$ 

Consider the region $x <0$. We will solve the equation for $\phi _ + ^ L (x)$ (the superscript L denotes the region $x <0$). Let's make a replacement:
\begin{equation}
    \xi = - \exp(-a(x+x_0)).
\end{equation}
Then the equation \eqref{eq:eq} takes the form:
\begin{equation}\label{eq:dirac1}
    0 =\left( a^2 \xi\frac{d}{d\xi}\left(\xi\frac{d}{d\xi}\right) - \left(m_f +\frac{A}{1-\xi}\right)^2 +  \frac{aA\xi}{(1-\xi)^2} + E^2 \right) \phi^L_{+}(\xi).
\end{equation}

The solution of the equation \eqref{eq:dirac1} is as follows:
\begin{equation}
\begin{aligned}
    \phi^L_{+}(\xi) &= C_1\xi^{-\alpha}(1-\xi)^{-\beta} \;_2F_1(-\alpha-\nu-\beta, -\alpha+\nu-\beta, 1-2\alpha;\xi)\\
    &+ C_2\xi^{\alpha}(1-\xi)^{-\beta} \;_2F_1(\alpha-\nu-\beta, \alpha+\nu-\beta, 1+2\alpha;\xi).
\end{aligned}    
\end{equation}
where the parameters $\alpha,\;\beta\;,\nu$ are defined as
\begin{equation}\label{eq:param}
\begin{aligned}
    &\alpha = \frac{1}{a}\sqrt{(m_f+A)^2 - E^2}= \frac{ip}{a}\\
    &\beta = -\frac{A}{a}  \\
    &\nu = \frac{1}{a}\sqrt{m_f^2-E^2} = \frac{i}{a}\sqrt{E^2 -m_f^2} = \frac{ik}{a}.
\end{aligned}    
\end{equation}

Let's consider the limit $x \rightarrow -\infty \Rightarrow \xi \rightarrow -\infty$. 
Then, for the function $\phi_+^L $ we obtain superposition of two waves: incident and reflected waves:
\begin{equation}\label{eq:asymp+}
    \phi_+^L \xrightarrow{x\rightarrow -\infty} D_{1}e^{ik(x+x_0)}+D_{2}e^{-ik(x+x_0)}.
\end{equation}
The coefficients $D_ {1},\;D_ {2}$ are determined by the formulas:
\begin{equation}\label{eq:coef}
\begin{aligned}
    D_{1} = C_1\frac{\Gamma(1-2\alpha)\Gamma(-2\nu)}{\Gamma(-\alpha-\nu-\beta)\Gamma(1-\alpha-\nu+\beta)}e^{-i\pi\alpha} +
    \\
    C_2\frac{\Gamma(1+2\alpha)\Gamma(-2\nu)}{\Gamma(\alpha-\beta - \nu)\Gamma(1+\alpha-\nu+\beta)}e^{i\pi \alpha}    
    \\
    D_{2} = C_1\frac{\Gamma(1-2\alpha)\Gamma(2\nu)}{\Gamma(-\alpha+\nu-\beta)\Gamma(1-\alpha+\nu+\beta)}e^{-i\pi\alpha} +
    \\
    +C_2\frac{\Gamma(1+2\alpha)\Gamma(2\nu)}{\Gamma(\alpha-\beta + \nu)\Gamma(1+\alpha-\nu+\beta)}e^{i\pi \alpha},
\end{aligned}
\end{equation}
where $C_1, C_2 = const$.
The asymptotic for  $\phi_-^L$ is obtained by substituting solution  $\phi_+^L$ \eqref{eq:asymp+}  into the first equation of system \eqref{eq:syst}. As a result, we obtain:
\begin{equation}
    \phi^L_-(x)\xrightarrow{x\rightarrow -\infty}-\frac{k+im}{E}D_1 e^{ik(x+x_0)} +
    \frac{k-im}{E}D_2 e^{-ik(x+x_0)}.
\end{equation}
The region $x>0$ can be considered in the similar manner to obtain the function  $\phi^R_+(x)$ to the right of the wall.

In order to find coupling between coefficients $C_1$ and $C_2$ we match the solutions at $x=0$:
\begin{equation}
\begin{aligned}
    &\phi^R_+|_{x=0} = \phi^L_+|_{x=0}\\
    &(\phi^R_+)'|_{x=0} = (\phi^L_+)'|_{x=0}.
\end{aligned}
\end{equation}

The normal component of the fermion current density is written as:
\begin{equation}\label{eq:curr1}
\begin{aligned}
    &j = \bar\psi (x) \gamma^3 \psi(x) = -|u_1(x)|^2+|u_2(x)|^2+|u_3(x)|^2-|u_4(x)|^2 =
    \\
    &=-\phi^*_+\phi_- -\phi^*_-\phi_+.
\end{aligned}
\end{equation}
Substitute the obtained functions As a result, we obtain
The final form of the current
\begin{equation}\label{eq:curr2}
    j = \frac{k}{E}(|D_1|^2-|D_2|^2) = j_{inc} - j_{ref},
\end{equation}
is obtained by substitution of the explicit form of functions $\phi ^ L _ +$ and $ \phi ^ L _-$ into the expressions for the current density. Here $j_ {inc}$ is the current density of the incident particles, $j_ {ref}$ - of the reflected ones. 

\begin{figure}[h!]\center
	\includegraphics[width=\textwidth]{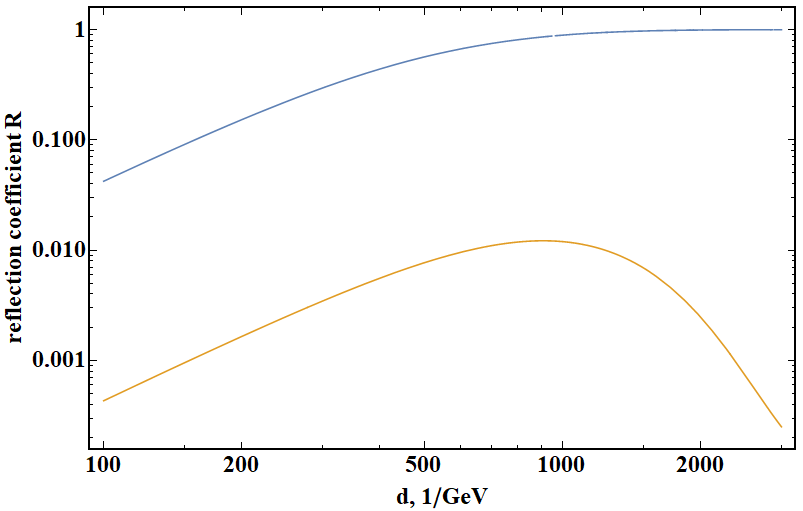}
		\caption{Dependence of the reflection coefficient R on thickness d, $GeV^{-1}$. Blue line - $m_0=10^{-3}\;GeV,\;E-m_f=1\;MeV$; orange line - $m_0=10^{-4}\;GeV,\;E-m_f=1\;MeV$}\label{reflection1}
	\includegraphics[width=\textwidth]{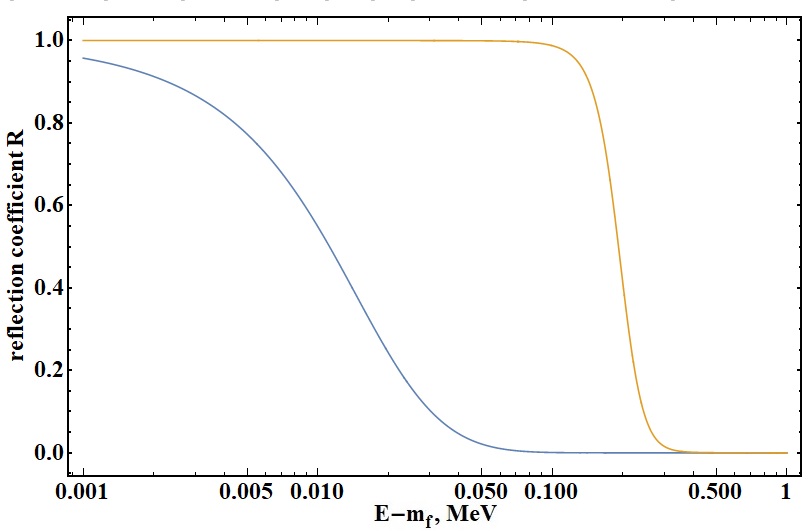}
		\caption{Dependence of the reflection coefficient R on kinetic energy $E-m_f\;MeV$. Blue line - $m_0=10^{-5}\;GeV,\;d=10^4\;GeV^{-1}$; orange line - $m_0=10^{-4}\;GeV,\;d=10^4\;GeV^{-1}$}\label{reflection2}
	\end{figure}

The reflection and transmission coefficients are determined through the ratio of the current densities as follows
\begin{equation}
    R = \frac{j_{ref}}{j_{inc}} = \frac{|D_2|^2}{|D_1|^2}.
\end{equation}
The coefficients $D_1,\;D_2$ are determined by the formulas \eqref{eq:coef}. The results of calculating the reflection coefficient for electrons ($m_f = 0.5\;MeV$) are shown in Figure \ref{reflection1},\ref{reflection2}.

\clearpage

\section{Conclusion}\label{s:conclusion}
The deceleration of the primordial walls due to the interaction with the surrounding media is the important process that could influence the formation of the black holes clusters. In this paper, we have found the reflection probability of the fermions. This is necessary step for studying the cluster heating by the wall fluctuation.

\section*{Acknowledgements}

This research was funded by the Ministry of Science and Higher Education of the Russian Federation, Project ``Fundamental properties of elementary particles and cosmology'' N 0723-2020-0041 and RFBR grant 19-02-00930.
The work of S.R. is performed according to the Russian Government Program of Competitive Growth of Kazan Federal University.


\printbibliography

\end{document}